\documentclass[oldversion,preprint]{aa}
\usepackage{graphicx}
\usepackage{txfonts}
\begin{document}
   \title{On the influence of the Sun on the rapid variability of compact extragalactic sources}
   \author{N.~Marchili \inst{1} \and
           T.~P.~Krichbaum \inst{1} \and
           X.~Liu \inst{2} \and
           H.-G.~Song \inst{2} \and
           J.~M.~Anderson \inst{1} \and
           A.~Witzel \inst{1} \and
           J.~A.~Zensus \inst{1}
          }

   \offprints{N. Marchili}

   \institute{Max-Planck-Institut f\"ur Radioastronomie, Auf dem H\"ugel 69,
              53121 Bonn, Germany\\
              \email{marchili@mpifr-bonn.mpg.de}
         \and
             Urumqi Observatory, the National Astronomical Observatories, the
             Chinese Academy of Sciences, Urumqi 830011, PR China
             }

   \date{Received ...; accepted ...}

\abstract{Starting from December 2004, a program for the monitoring of
  intraday variable sources at a frequency of 5\,GHz was performed at the
  Urumqi Observatory. The analysis of the variability
  characteristics of the flat-spectrum radio source AO 0235+164
  revealed the existence of an annual cycle in the variability
  amplitude. This appears to correlate with the solar elongation of the
  source. A thorough analysis of the results of the MASIV IDV survey ---
  which provides the variability characteristics of a large sample of compact
  radio sources --- confirms that there is a small but detectable component of the
    observed fractional modulation which increases with decreasing
    solar elongation. We discuss the hypothesis that the phenomenon is related
    to interplanetary scintillation.}


\authorrunning{N. Marchili et al.}
\titlerunning{}

   \maketitle
%
\section{Introduction}

IntraDay Variability (IDV, see Witzel et al. \cite{Witzel1986}, Heeschen et
al. \cite{Heeschen1987}) refers to
the fast intensity variability --- on time scales from a few hours to $\sim2$\,days ---
which affects a considerable number of flat-spectrum radio sources (a
fraction between  30\% and 50\%, see Quirrenbach et al. \cite{Quirrenbach1992} and
Lovell et al. \cite{Lovell2008}). The variability concerns both total flux density and
polarization measurements over a wide range of the electromagnetic spectrum,
from the radio to the optical bands. 

Both source-intrinsic and source-extrinsic models have been proposed in order
to explain IDV. In the optical bands, the variability should be regarded as
intrinsic to the sources (Wagner \& Witzel \cite{Wagner1995}). Concerning the
radio bands, the problem of the origin of the variability is not yet
solved. For the most extreme sources --- the so-called fast-scintillators (see
Dennett-Thorpe \& de Bruyn \cite{Dennett-Thorpe2002}, Bignall et
al. \cite{Bignall2003}), which show total flux variations of the order of
100\% on time scales of hours --- the variability is most likely due to InterStellar
Scintillation (ISS), caused by nearby scattering screens located at a distance of
several parsecs (see Dennett-Thorpe \& de Bruyn
\cite{Dennett-Thorpe2002}, Rickett et al. \cite{Rickett2006}). However,
only a very small fraction of IDV sources can be labeled as fast
scintillators.  In the Micro-Arcsecond Scintillation-Induced
Variability (MASIV) survey (Lovell et al. \cite{Lovell2003} and
\cite{Lovell2008}), a strong correlation was found between the
variability amplitude of a large sample of compact radio sources and
the emission measure in the ionized interstellar medium along their
respective lines of sight, showing that a significant part of the
variability is due to ISS.

The discovery of correlated variations in the optical and radio light curves
of S5 0716+714 (Quirrenbach et al. \cite{Quirrenbach1991}) suggested that
at least part of its variability is intrinsic to the source. This
gives rise to the fundamental issue to establish how large the
contribution of source-intrinsic mechanisms to the total variability
is and, even more important, whether the case of S5 0716+714 should be
regarded as an exception among IDV sources. 

A possible way to investigate the nature of IDV in a given source is to
study how its variability characteristics, namely amplitude and time
scale, change with time. Following Narayan \cite{Narayan1992}, we shortly
discuss how these quantities vary in case of IDV which is caused by ISS. They can
both be described in terms of the Fresnel scale, $r_{\mathrm{f}}$,
and the diffraction scale, $r_{\mathrm{diff}}$. The former is given by
$r_{\mathrm{f}}\sim \sqrt{\lambda\,D}$, where $D$ is the distance between the
observer and the scattering screen. The latter depends on the properties of
the screen and the wavelength of the observations; concerning the
  interstellar medium, typical
values of $r_{\mathrm{f}}$ are of the order of $10^{10}$\,cm for
observations at centimeter wavelengths and screen distances of the
order of 10-100 pc. In case the variability is caused by
weak scattering --- as one would expect for classical IDV radio sources --- and the
angular source size $\theta_{\mathrm{s}}$ is larger than the Fresnel angle
$\theta_{\mathrm{f}}=r_{\mathrm{f}}/D$, the characteristic variability time
scale can be expressed as 

\begin{equation}
\tau_{\mathrm{c}} \approx \frac{r_{\mathrm{f}}}{v}\frac{\theta_{\mathrm{s}}}{\theta_{\mathrm{f}}}
\label{eq:ts}
\end{equation}
where $v$ is the relative velocity between the screen and the observer. 

For a given time series, a measure of its variability amplitude is given by the
modulation index, $m_{\mathrm{i}}$, which is the ratio of the standard
deviation to the average. In case of ISS-originated variability, the 
modulation index of a light curve is given by Narayan (\cite{Narayan1992}) as

\begin{equation}
m_{\mathrm{i}} \approx \Big(\frac{r_{\mathrm{f}}}{r_{\mathrm{diff}}}\Big)^{5/6}\Big(\frac{\theta_{\mathrm{f}}}{\theta_{\mathrm{s}}}\Big)^{7/6}
\label{eq:mind}
\end{equation}

The two equations above can be used to predict how the variability
characteristics of a source should evolve throughout the year. The time scale
of scintillation-induced variability changes with the relative velocity
between the scattering screen and the observer. Due to the Earth's motion
around the Sun, this velocity follows an annual cycle which should result in
an annual modulation of the variability time scale (see Dennett-Thorpe \& de
Bruyn \cite{Dennett-Thorpe2000}, Rickett et al. \cite{Rickett2001}, Gab\'anyi
et al. \cite{Gabanyi2007}). The
modulation index, instead, should not change with time, unless the variability
time scale exceeds the total duration of the observations, in which case
the detected variability would decrease as  the time scale increases.  

\begin{figure}
   \centering
   \includegraphics[width=0.96\columnwidth]{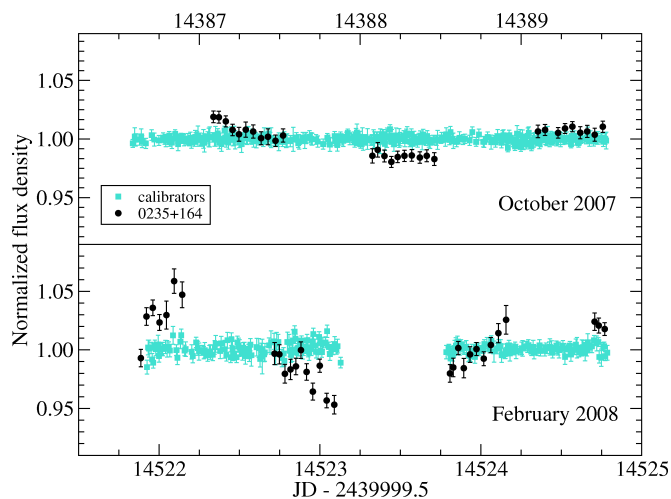}
      \caption{
        The variability curves of AO 0235+164 (black dots) in
        October 2007 (upper panel) and February 2008 (lower panel),
        examples, respectively, of a low and a high variability
        state of the source.}   
      \label{fig:lc3}
\end{figure}

In 2004, a collaboration between the Max-Planck-Institut f\"ur Radioastronomie
(MPIfR) and the Urumqi Observatory initiated a project for the monitoring of
classical IDV sources (Gab\'anyi et al. \cite{Gabanyi2007}, Marchili et
al. \cite{Marchili2008} and \cite{Marchili2010}). The aim
of the project is to study how the variability characteristics of the target
sources change with time. Among others, a sample of 6 well known IDV sources
has been regularly observed with the 25m-Urumqi radio telescope, at a
frequency of 4.80\,GHz. These are the flat-spectrum radio sources
AO 0235+164, 0716+714, 0917+624, 0954+658, 1128+592 and 1156+295. The project is
still ongoing; through February 2010, 42 observing sessions were carried
out. The main characteristics of the 21 epochs in which
  AO 0235+164 was observed are summarized in Table \ref{tab:mindex}: in
Col.\,1 we report the observing date at half-session, in Col.\,2 the
corresponding day (0=January 1st, 2005), in Col.\,3 the duration of
the observations, in Col.\,4 the number of observed sources (including
the calibrators) and in Col.\,5 the duty cycle (i.e. the average
number of data-points per source per hour). 

\section{Urumqi data}
\subsection{Observation and data calibration}

The observations have been performed with the 25-meter parabolic antenna of
the Nanshan radio telescope, operated by the Urumqi Observatory (for more
details, see Sun et al. \cite{Sun2006}, Marchili et al. \cite{Marchili2010}
and references therein). Its single beam dual polarization receiver, built by
the MPIfR, is centered at a frequency of 4.80\,GHz and has a bandwidth of
600\,MHz. 

\begin{figure}[htbp!]
   \centering
   \includegraphics[width=0.96\columnwidth]{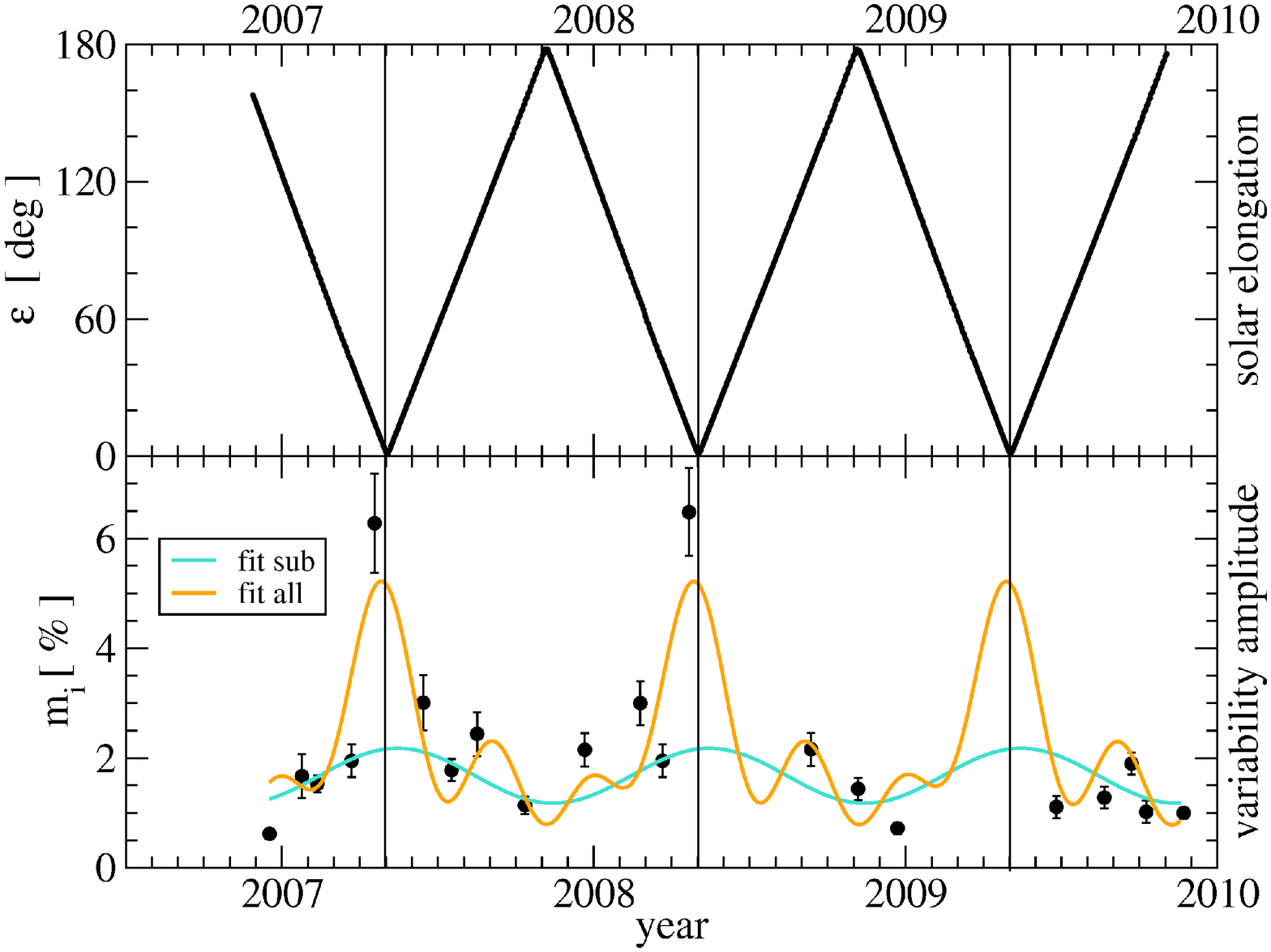}
      \caption{Solar elongation (upper panel) and modulation
        index (lower panel) of AO 0235+164 plotted versus observing
        time. The orange line shows the sum of the three variability
        components which correspond to the harmonics of the
        signal highlighted in the periodogram analysis in
        Fig.\,\ref{fig:prd}. The cyan line shows a sinusoidal fit
        obtained after removing the data-points corresponding to the
        two epochs at smallest solar elongation.
              }
              \label{fig:mind1}
\end{figure}

All the flux density measurements have been performed in cross-scan
mode, each scan consisting of 8 sub-scans --- 4 in azimuth, 4 in elevation ---
over the source position. This observing mode allows the evaluation and
correction of residual small pointing offsets and the detection of
non-Gaussian cross-scan profiles in case of in-beam confusion.

The data calibration procedure follows a standard pipeline. A Gaussian fit to
the sub-scans provides an estimate of the flux density. After a quality check,
an error-weighted average on the reliable sub-scans provides a flux density
measurement for each scan . This value is then corrected for the antenna-gain
dependency on both the elevation and the weather conditions, by parameterizing
the changes that these effects induce on several calibrators. The accuracy in
the flux density measurements can be evaluated through the modulation index of
the calibrators, $m_{\mathrm{0}}$. During normal weather conditions, we find
$m_{\mathrm{0}}$ values between 0.5 and 0.7\%. A more thorough description of
the data calibration procedure can be found in Kraus (\cite{Kraus1997}),
Marchili (\cite{Marchili2009})  and references therein. 

\subsection{Variability characteristics of AO~0235+164}

The flat-spectrum radio source  AO 0235+164 (z=0.94) was found to
  show variability on IDV time scales more than once in the past (see,
  e.g., Kraus et al. \cite{Kraus1999}); Senkbeil et
  at. (\cite{Senkbeil2008}) reported a likely
  extreme-scattering-event in AO 0235+164 in July 2005.  The source
was observed with the Urumqi radio telescope in 21 epochs between
December 2006 and November 2009. Two of the collected light
  curves are shown in Fig.\,\ref{fig:lc3}. In Table \ref{tab:mindex},
for each observing session we summarize the calibration accuracy
($m_{\mathrm{0}}$; Col.\,6) and the variability amplitude of AO 0235+164
(Col.\,7) in terms of the modulation index $m_{\mathrm{i}}$. The uncertainties on  $m_{\mathrm{i}}$ --- evaluated by means of synthetic light curves having the same sampling as the original curves --- range between 10 and 20\% of the $m_{\mathrm{i}}$ estimations.

\begin{table}[ht]
  \caption{The observing sessions of the Urumqi monitoring program
    in which AO 0235+164 was observed.}  
  \label{tab:mindex}  
  \centering                    
  \begin{tabular}{l c c c c c c} 
    \hline\hline   
Epoch & Day &  Duration & N. S. & Duty cycle &  $m_{\mathrm{0}}$ & $m_{\mathrm{i}}$\\
      &     &    (d)    &            & (data\,h$^{-1}$)        &       (\%) & (\%)    \\
    \hline      
2006.12.18 &   718 & 2.4 & 12 & 1.3 &  0.6 & 0.62 \\
2007.01.25 &   755 & 2.3 & 14 & 1.1 &  0.7 & 1.67 \\
2007.02.12 &   773 & 4.0 & 15 & 1.0 &  0.6 & 1.53 \\
2007.03.24 &   813 & 2.8 & 16 & 0.9 &  0.7 & 1.95 \\
2007.04.20 &   840 & 3.7 & 16 & 0.8 &  0.8 & 6.28 \\ 
2007.06.16 &   897 & 2.4 & 16 & 0.9 &  0.7 & 3.01 \\ 
2007.07.19 &   930 & 2.9 & 18 & 0.9 &  0.7 & 1.78 \\ 
2007.08.18 &   960 & 3.1 & 15 & 1.0 &  0.8 & 2.44 \\ 
2007.10.13 &  1016 & 3.0 & 16 & 0.8 &  0.5 & 1.14 \\
2007.12.22 &  1086 & 3.2 & 15 & 1.0 &  0.5 & 2.15 \\
2008.02.25 &  1151 & 2.9 & 15 & 0.8 &  0.6 & 3.00 \\
2008.03.22 &  1177 & 3.0 & 15 & 1.1 &  0.5 & 1.95 \\
2008.04.22 &  1208 & 3.1 & 14 & 0.9 &  0.5 & 6.48 \\ 
2008.09.12 &  1351 & 3.5 & 14 & 1.1 &  0.5 & 2.16 \\
2008.11.06 &  1406 & 3.6 & 15 & 0.5 &  0.7 & 1.44 \\
2008.12.22 &  1452 & 2.4 & 15 & 1.0 &  0.5 & 0.72 \\
2009.06.26 &  1638 & 2.6 & 16 & 0.7 &  0.5 & 1.11 \\ 
2009.08.21 &  1694 & 4.1 & 16 & 0.9 &  0.5 & 1.28 \\
2009.09.22 &  1726 & 5.5 & 14 & 0.9 &  0.5 & 1.90 \\
2009.10.09 &  1743 & 2.3 & 15 & 1.1 &  0.5 & 1.02 \\
2009.11.22 &  1787 & 3.8 & 16 & 0.8 &  0.6 & 1.00 \\
\hline
  \end{tabular}
\end{table}

\subsection{Annual variation in the modulation index of AO~0235+164}

\begin{figure}
   \centering
   \includegraphics[width=0.96\columnwidth]{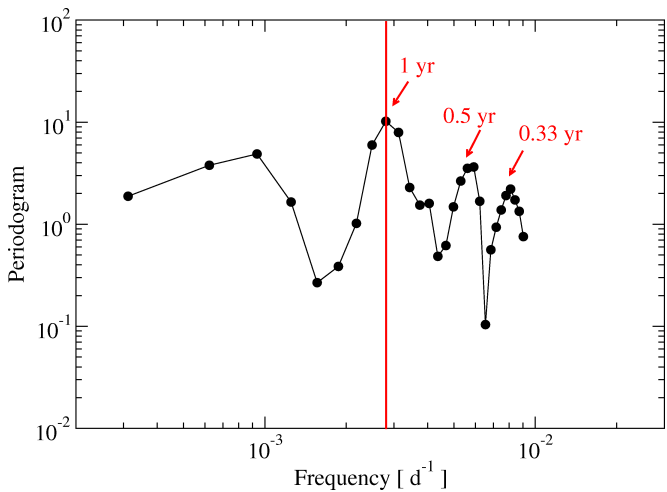}
      \caption{Periodogram analysis of the variability amplitudes
        $m_{\mathrm{i}}$ of AO 0235+164. Four peaks of power are clearly
        visible; three of them correspond to periods of 1, 1/2 and 1/3 of
        year (red arrows), i.e. the first three harmonics of a 1-year periodic
        signal. This is strong evidence in favour of an annual cycle in the
        variability amplitude of the source.
              }
              \label{fig:prd}
\end{figure}

Plotting $m_{\mathrm{i}}$ versus the date of the observation, it appears that the
modulation index of AO 0235+164 (black dots in Fig. \ref{fig:mind1}, lower
panel) follows a regular pattern. The variability seems to be more intense
between February and August, weaker between September and January.
To investigate this effect, we applied  to $m_{\mathrm{i}}$ a
Lomb-Scargle periodogram analysis (see Lomb \cite{Lomb1976}, Scargle
\cite{Scargle1982}), which highlighted the existence of a periodic
oscillation with a period of one year (see Fig.\, \ref{fig:prd}).

We summed up the variability components corresponding to the
three harmonics of the signal which are indicated in the periodogram analysis
in Fig.\,\ref{fig:prd}. This allows an estimate of
the amplitude and phase of the periodic oscillation. The annual cycle (orange
line in Fig. \ref{fig:mind1}, lower panel) peaks at the time of the year
in which the solar elongation of AO 0235+164, $\epsilon_{0235}$, is at the minimum
(see Fig. \ref{fig:mind1}, upper panel). The yearly variation observed in
$m_{\mathrm{i}}$ has a high amplitude in 2007 and 2008. In 2009, the
modulation is not clearly visible. This may be due to the lack of observations
between January and July, which is the time span in which we would expect the
largest variations. However, we may notice that between 2008 and 2009 the
solar activity was very low, due to the transition from the 23rd to the 24th
solar cycle. In this sense, the absence of significant changes in the
variability amplitude of AO 0235+164 may support the hypothesis that these
changes are related to the solar activity and therefore are induced by the
Sun. 

Two measurements of $m_{\mathrm{i}}$ stand out from the others; they correspond to the
observing sessions performed in April 2007 and April 2008. In those
epochs, the solar elongation 
$\epsilon_{0235}$ was relatively small ($\sim13^\circ$ and $\sim11^\circ$,
respectively). 
We repeated the periodogram analysis excluding the two data-points with
$\epsilon_{0235}<30^\circ$, in order to establish if the increase of the $m_{\mathrm{i}}$
value is limited to the sole epochs of small solar elongation. We found that a
periodic oscillation with period of about one year is still clearly
detectable. A sinusoidal fit to these data (see Fig. \ref{fig:mind1}, lower
panel, cyan line) shows that the oscillation has a peak-to-peak
amplitude of $\sim1.1$\%, while the average $m_{\mathrm{i}}$ value is
$\sim1.7$\%. The increase in $m_{\mathrm{i}}$ between the time of maximum
and minimum solar elongation is of the order of 90\%. This implies
that, during our observations, the solar
elongation plays the main role in the variability of AO 0235+164. The
phase of the periodic signal  
is slightly offset ($\sim10$ days) with respect to the time of the year of
minimum elongation. This offset, however, is much smaller than the average
time separation between consecutive observing sessions ($\sim$ 50 days). We
can conclude that periodic variations correlated to solar elongation
affect the modulation index of AO 0235+164 even for solar elongation
$\epsilon_{0235}$ larger than $30^\circ$. 

\subsection{Variability time scale}
\label{sec:vts}

At first sight, one may think that the existence of an annual cycle in the
modulation index of an IDV source may be related to an annual modulation in
its variability time scales. As explained above, $m_{\mathrm{i}}$ could undergo
yearly-periodic changes if the variability time scale exceeds the duration of
the observations. 
In this case, the variations in $\tau_{\mathrm{c}}$ and $m_{\mathrm{i}}$ should
be anti-correlated.
We can check this hypothesis by studying how
$\tau_{\mathrm{c}}$ changes as a function of $m_{\mathrm{i}}$. For
each observing session, we estimated the characteristic
variability time scale of AO 0235+164  by applying to the
  light curves three different kinds of time analysis methods, namely
  a first-order structure function analysis (see Simonetti et
  al. \cite{Simonetti1985}), a wavelet-based algorithm (Marchili et
  al., in prep.) and a sinusoidal fitting procedure. 
The uncertainties in the time scales are mostly due to the limited
duration of the observations, $obs_d$. We estimated them as proportional to
$({\tau_{\mathrm{c}}}^{3/2})/{(obs_d)}^{1/2}$. The
proportionality factor has been calculated by looking at the
distribution of the peak-to-peak time scale --- the time interval
between a local minimum (or maximum) and the following maximum (or
minimum) --- for a few light curves characterized by fast variability.


\begin{figure}
   \centering
   \includegraphics[width=0.96\columnwidth]{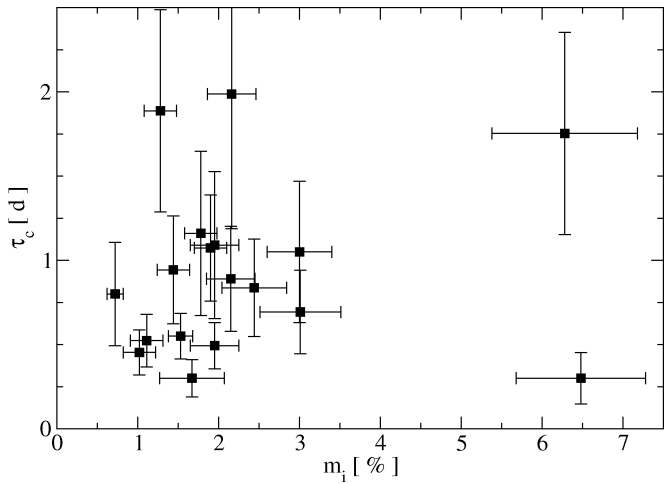}
      \caption{The variability time scales of AO 0235+164 plotted versus $m_{\mathrm{i}}$.
              }
              \label{fig:ts}
\end{figure}

\begin{figure}
   \centering
   \includegraphics[width=0.96\columnwidth]{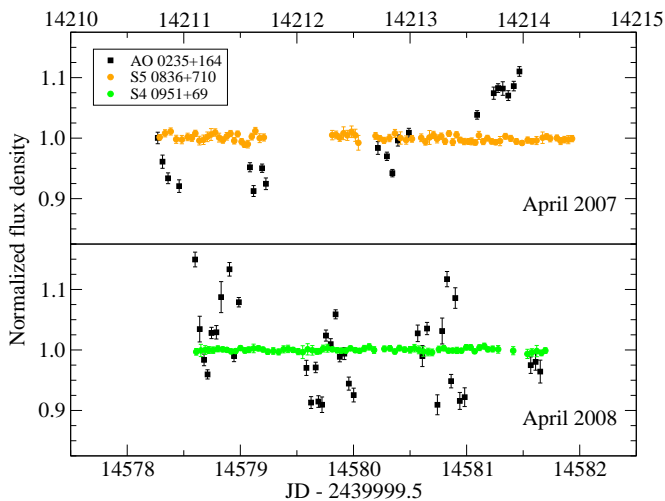}
      \caption{The variability curves of AO 0235+164 (black squares) in April 2007 (upper
        panel) and April 2008 (lower panel), compared with the curves of the calibrators
        S5~0836+710 (orange dots) and S4~0951+69 (green dots). The time scale of
        the variability in the two epochs is considerably different.} 
      \label{fig:lc}
\end{figure}

In Fig. \ref{fig:ts} we plot the variability time scales versus
$m_{\mathrm{i}}$ for all the epochs in which the source showed significant
variability (i.e. the probability of constant flux density was lower than
0.1\%, according to a chi-square test).  A cycle in $m_{\mathrm{i}}$ as strong
as the one we observed in AO 0235+164 should result in a clear increase of
$\tau_c$ as $m_{\mathrm{i}}$ decreases. The plot, however, does not reveal
such a trend. In particular, while the extreme variability observed in April 2007 is characterized by a value of $\tau_{\mathrm{c}}$ among the highest detected in the source, the  one for the April 2008 observations is quite low (see Fig.\,\ref{fig:lc}). This
leads to the conclusion that the annual cycle in $m_{\mathrm{i}}$ cannot be
explained in terms of an annual modulation of the variability time scale. 

\section{MASIV data}

Given the large
number of sources it comprises, the MASIV survey seems to be the ideal
test ground to further investigate the possible correlation
  between the variability characteristics of compact radio sources and
  their solar elongation.

MASIV is a survey of 710 radio sources, undertaken at a frequency of 4.9\,GHz
at the Very Large Array (VLA) between January 2002 and January
2003. The main aim of the project was to provide a large sample of
scintillating sources for reliable statistical investigation. A core
sample of 578 sources was observed in four epochs of 3 or 4 days
duration, starting on 2002 January 19, May 9, September 13 and 2003
January 10. The results of these observations, along with a
detailed description of the observing strategy and data calibration,
are reported in Lovell et al. (\cite{Lovell2008}). After removal of
the sources which show structure on VLA arcsecond scales or are partially
resolved, 475 point sources were left. Basic information about the
sources, their flux density and the raw modulation index for each of
the four epochs are provided by Lovell et al. (\cite{Lovell2008}), and
are available in the electronic edition of the {\it Astrophysical Journal}. 

 We used the modulation indices
resulting from the MASIV survey to check whether the data support
  the hypothesis of an additional contribution to the variability,
  related to solar elongation. We also checked whether such an 
  effect depends on the ecliptic latitude of the sources, as it would
  be reasonable to expect. 

\subsection{Modulation index variations as a function of solar elongation}

We labeled the four epochs of MASIV observation chronologically, as t$_j$
\{j=1,..., 4\}. For each session and source, we calculated the solar elongation $\epsilon_{src,
  j}$. We also defined a new parameter, the fractional modulation index, as

\begin{equation}
fm_{\mathrm{src,\ j}}=\frac{m_{\mathrm{src,\ j}}}{<m_{\mathrm{src}}>}
\label{eq:fm}
\end{equation}
where $m_{\mathrm{src,\ j}}$ is the modulation index at the
epoch t$_\mathrm{j}$ and $<m_{\mathrm{src}}>$ is the average modulation index over
the four observing sessions. 
By means of the fractional modulation index, we can compare the
  changes in the variability amplitude of sources with very different
  variability characteristics. In a statistical approach, the
 $fm$ values have been combined, in order to study how the
variability amplitude changes, on average, as a function of
$\epsilon$. 
We excluded from the analysis the light curves which
  have been used to calibrate the data (H.~Bignall, priv. comm.), because 
  their modulation index is artificially low.

\begin{figure}
   \centering
   \includegraphics[width=0.96\columnwidth]{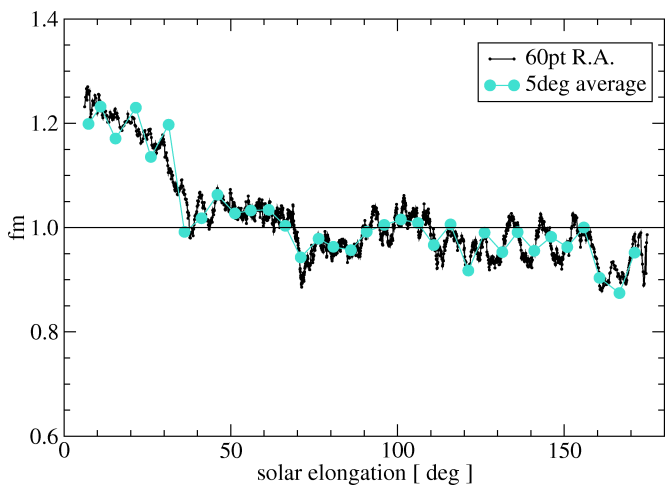}
      \caption{Results of a 60-point running average on the
        combined fractional modulation index of MASIV sources (black
        line) plotted versus solar elongation. The cyan dots show a
        5-degree average. The increase in the variability amplitude 
        for $\epsilon < 30^\circ$ is remarkable.
              }
              \label{fig:fmall}
\end{figure}

If the effect observed in AO 0235+164, described above, is
common to compact extragalactic sources, we expect to see an
increase in $fm$ as $\epsilon$ approaches zero. At first sight, the scattering
in the individual $fm$ values is too high to reveal a clear trend. 
A 60-point running average over the data (see Fig. \ref{fig:fmall}, black line)
demonstrates the existence of such an increase. In cyan dots, we also
plotted the 5-degree average, which confirms the 
  behaviour of the running average.  The averaged $fm$ peaks at
$\epsilon\sim 10^\circ$ with a value of $\sim 1.25$, while the
minimum falls close to $\epsilon \sim165^\circ$ with a value of
$\sim  0.9$. Most remarkably,
all the averaged $fm$ values for $\epsilon<35^\circ$ are $\gtrsim 1.15$, while
for $\epsilon> 110^\circ$ most of the values are $< 1.0$.
We conclude that solar
elongation exerts a considerable influence on the variability of compact radio
sources. 

 Concerning the MASIV survey, the contribution of the
  solar-elongation related effect to the total variability should  not
  have very serious implications for its main findings. In particular,
  all the results obtained by using the structure function at a time
  lag of two days ($D$(2 days) in Lovell et al. \cite{Lovell2008}) as
  an estimate of the source variability should be only marginally
  affected. This is because $D$(2 days) is calculated over all the
  four observing sessions. Based on Fig. \ref{fig:fmall}, given a
  30-40\% enhancement in the modulation index between maximum and
  minimum solar elongation, the additional contribution from the solar
  elongation effect to the 4-epoch combined variability would be of
  the order of 10\% of the total fractional variation for
  sources observed at low solar elongations. 

\subsection{Dependence on ecliptic latitude}

A solar elongation related effect should be particularly strong for sources 
at low ecliptic latitude, due to the large solar elongation range they
cover. Let us define $\epsilon_{min}$ and $\epsilon_{max}$ as the minimum and
maximum solar elongation for a given source during the year. We would expect 
$m_{\mathrm{i}}(\epsilon_{min})-m_{\mathrm{i}}(\epsilon_{max})$ (from now on,
$\Delta m_{\mathrm{max}}$) to increase as the ecliptic latitude $\beta$ tends
to zero. In order to verify the hypothesis, for each source in the MASIV 
catalogue we applied a linear regression to the four $m_{\mathrm{i}}$ values
as a function of $\epsilon$. The regression coefficient $A$ has been
used to estimate the variation in the modulation index, as follows: 

\begin{equation}
\Delta m_{\mathrm{max}}=A\,(\epsilon_{min}-\epsilon_{max})
\end{equation}

\begin{figure}
   \centering
  \includegraphics[width=0.96\columnwidth]{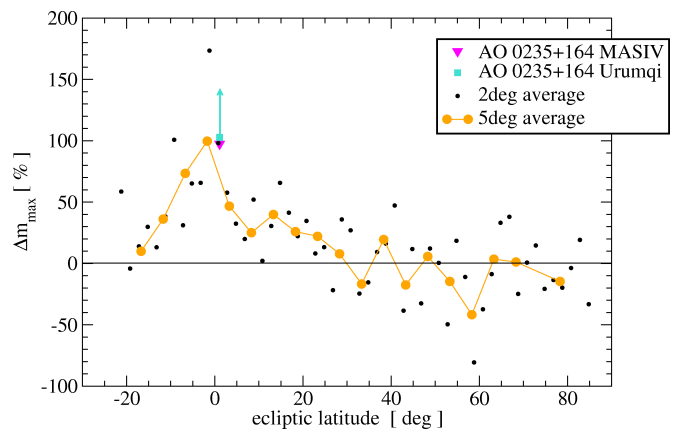}
      \caption{The modulation index change, $\Delta m_{\mathrm{i}}$, as a source passes
        from the farthest to the closest point to the Sun plotted
      versus ecliptic latitude $\beta$. The 2-degree (black dots) and
      5-degree average (orange dots) pinpoint the large increase in variability
      for sources with low $\beta$. The variation deduced from Urumqi
      observation for AO 0235+164, plotted as a cyan square, has been
      obtained excluding the April 2007 and April 2008 observations, and
      should be regarded as a lower limit.
      }
              \label{fig:beta}
\end{figure}

The $\Delta m_{\mathrm{max}}$ values obtained for all sources, averaged in
bins of 2 (black dots) and 5 degrees (orange dots), are
presented in Fig.\,\ref{fig:beta}. $\Delta m_{\mathrm{max}}$ is clearly
$\beta$-dependent. For sources at $\beta\sim0^\circ$, $\Delta 
m_{\mathrm{max}}$ is of the order of 100\%. It decreases on both sides of the $\beta$
axis, till it reaches a value close to 0\% at $\beta\sim30^\circ$.

 The calculation of the mean fractional modulation index values at
different epochs reveals an interesting characteristic. The averages for the
January 2002 and 2003 epochs are systematically smaller than for May and
September 2002. The measured values are shown in Table
\ref{tab:fmepochs} (Col.\,2). This result needs to be carefully examined. The
distribution of the sources as a function of solar elongation changes during
the year. In January the Sun is at low declination. Since the MASIV
survey includes only sources above declination 0$^{\circ}$, in January
the average solar elongation is considerably higher than
in May and September. As a consequence, a systematic bias in the measured
modulation indices at different epochs --- whatever the cause --- could in
principle fake a solar elongation dependence of the
variability. Viceversa, a solar elongation dependence of the
variability would affect the average modulation indices measured at
different times of the year. If the first hypothesis is correct, the
average values of $fm$ should not depend on $\beta$. If the second
hypothesis is correct, the differences in $fm$ at different epochs
should be very pronounced for sources at low $\beta$, negligible for
the ones at high $\beta$. 

\begin{table}[t]
  \caption{The mean fractional modulation index of the four MASIV observing
    sessions. The average fractional modulation index is compared with
    the averages for the subsamples of source with
    $|\beta|<15^\circ$ and $|\beta|>45^\circ$.}  
  \label{tab:fmepochs}  
  \centering                    
  \begin{tabular}{l c c c} 
    \hline\hline   
           Epoch & $\bar{fm}$  &  $\bar{fm}$($|\beta|<15^\circ$) &  $\bar{fm}$($|\beta|>45^\circ$) \\
    \hline
  January 2002   & 0.92      &   0.85             &   1.04 \\
      May 2002   & 1.10      &   1.18             &   1.03 \\
September 2002   & 1.11      &   1.17             &   1.05 \\
  January 2003   & 0.87      &   0.80             &   0.88 \\
    \hline
  \end{tabular}
\end{table}

The $fm$ averages for two subsamples of sources at low ($<15^\circ$) and high
($>45^\circ$) ecliptic latitude are reported in Table \ref{tab:fmepochs},
Col.\,3 and 4 respectively. 

It appears that the fractional modulation index is strongly
$\beta$-dependent even within a single observing session, which
confirms the existence of a solar-elongation related effect. The similarity
between the averages at high $\beta$ for January, May, and September 2002 seem
to indicate that the low average $fm$ value in January 2002 is mainly due to
solar elongation. In the case of January 2003, instead, the low $fm$ values
at both low and high $\beta$ cannot be solely ascribed to a solar elongation
effect. 

 We
  compared the $fm$ averages in Col. 2 of Table  \ref{tab:fmepochs}
  with the number of variable sources, N$_\mathrm{var}$, detected in
  each epoch, reported in  Fig.\,5 of Lovell et
  al. \cite{Lovell2008}. N$_\mathrm{var}$  increases by $\sim 20\%$
  from January 2002 to May,  by a few percent from May to September and
  decreases of $\sim 25-30\%$ from September to January 2003. This is
  in excellent agreement with the behaviour of the $fm$ averages,
  confirming that the solar-elongation related effect significantly
  contributes to the variability observed in MASIV sources.

\subsection{Comparison between MASIV and Urumqi results}
\label{comp}

We compared the results obtained from the MASIV sample with the ones from
the Urumqi observations. AO 0235+164 has an ecliptic latitude
$\beta\sim1^\circ$; it was observed both in the MASIV survey and in the
Urumqi IDV monitoring project. In the first case, we find a $\Delta
m_{\mathrm{max}}$ value of 97\% (plotted in Fig.\,\ref{fig:beta} as a
purple triangle), in excellent agreement with the sources-averaged $\Delta
m_{\mathrm{max}}$ at $\beta \sim 0^\circ$, which is 98\%. If we calculate in a similar
way the $\Delta m_{\mathrm{max}}$ for the Urumqi observations (cyan
square in Fig.\,\ref{fig:beta}), we find values which range from $\sim 90$ to
almost 400\%, depending on whether the April 2007 and 2008 epochs are excluded
from the calculation or not. The exceptional increase in the
variability observed in Urumqi for $\epsilon\sim10^\circ$ finds no
confirmation in the observations of AO 0235+164 in the May epoch
  of MASIV, when the solar elongation of the source was below $10^\circ$. 

\section{Discussion}
\label{Discussion}

The combined analysis of Urumqi and MASIV results has provided enough evidence
to claim the existence of a source of variability which is related to solar
elongation. The nature of the variability, however, is not yet
clear. Below, some possible explanations are proposed. The discussion 
of the different models is based on the results of the Urumqi observations, for which 
we could estimate both the amplitude and the time scales of the variability.

\subsection{Weak scattering}

Assuming that the variability is caused by weak scattering, we can use
Eq.\,\ref{eq:ts} and \ref{eq:mind} to figure out what conditions would be
needed to match the variability characteristics we observed --- namely, $m_{\mathrm{i}}
\sim 1$--$10\%$ and $\tau_{\mathrm{c}} \sim 10^4$--$10^5$\,s. Assuming a scattering screen at
1\,a.u. distance, we can calculate a Fresnel scale $r_{\mathrm{f}} \sim 10^2$\,km and a
Fresnel angle $\theta_{\mathrm{f}} \sim 10^2$\,mas. The only parameters we can modify to
obtain the proper time scale are the angular size of the emitting region
$\theta_{\mathrm{s}}$ and the relative velocity between the screen and the observer,
$v$. 

Setting a velocity of the order of the solar wind speed
($\sim10^2$\,km\,s$^{-1}$) would lead to $\theta_{\mathrm{s}} \geq
10^4\,\theta_{\mathrm{f}}$, which means $100''$ at least. Looking at
the full-width at half maximum of cross-scans obtained with the Urumqi and
Effelsberg radio telescopes, we can rule out this hypothesis.

Assuming $\theta_{\mathrm{s}}$ to be of the order of milliarcseconds or smaller (see, e.g., Lazio et 
al. \cite{Lazio2008} and Qian \& Zhang \cite{Qian2001}), the source must be regarded as pointlike, and the
factor $\theta_{\mathrm{s}}/\theta_{\mathrm{f}}$ in Eq.\,\ref{eq:ts} and
\ref{eq:mind} can be substituted with 1. The constraint on the
variability time scale would require $v \leq 10^{-2}$\,km\,s$^{-1}$. This is three orders of magnitude lower than the Earth's orbital velocity, which, reasonably, should provide a lower limit to $v$.

Looking at the literature, it is well known that InterPlanetary Scintillation (IPS)
does contribute to the variability of compact radio sources (Readhead
\cite{Readhead1971}). IPS causes variations of the scintillation index (the
standard deviation of the flux density over an ensemble of measurements; it is
equivalent to $m_{\mathrm{i}}$) 
which depend on solar elongation. The effect is most prominent at low observing
frequencies. Following Readhead \cite{Readhead1971}, for $\epsilon$
between $\sim 3^\circ$ and $\sim 90^\circ$, and for frequencies between 81.5\,MHz
and 2.7\,GHz, in the regime of weak scattering, the dependence of
$m_{\mathrm{i}}$ on $\sin \epsilon$ and the frequency $\nu$ can be
described as a power law  

\begin{equation}
m_{\mathrm{i}}(\nu)\,\nu \propto (\sin\,\epsilon)^{-1.55}
\end{equation}
Extrapolating this result to a frequency of 4.8\,GHz, and using a
proportionality factor of 0.22 (see Readhead \cite{Readhead1971})
gives for sources with $\epsilon<30^\circ$ a $m_{\mathrm{i}}$ of the order  of 1-10\%,
which is consistent with  our results. The time scale of IPS variations,
however, is of the order of seconds, at least $10^4$ times smaller
than the ones which characterize the Urumqi light curves.

\subsection{Strong scattering: refractive scintillation}

\begin{figure}
   \centering
  \includegraphics[width=0.96\columnwidth]{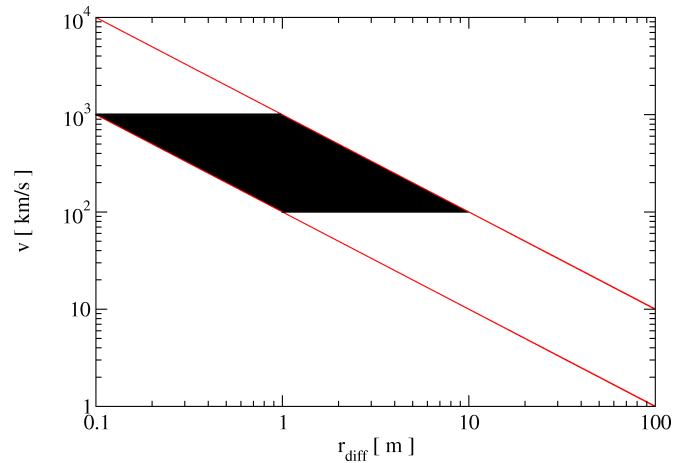}
      \caption{Between the red lines, the range of values $r_{\mathrm{diff}}$ (x-axis) and
        $v$ (y-axis) which are consistent with the variability characteristics 
        observed in the Urumqi data. In black, the subset of values in the
        $v$-range $10^2$--$10^3\,$km/s, corresponding to the solar wind speed.
      }
              \label{fig:refpar}
\end{figure}

Let us hypothesize that the variability is due to refractive
scintillation. In the case of extended sources, the time scale and modulation
index of the variability would be 

\begin{equation}
\tau_{\mathrm{ref}} \approx \frac{r_{\mathrm{ref}}}{v}\frac{\theta_{\mathrm{s}}}{\theta_{\mathrm{scatt}}}
\label{eq:tsref}
\end{equation}
and

\begin{equation}
m_{\mathrm{i, ref}} \approx \Big(\frac{r_{\mathrm{diff}}}{r_{\mathrm{f}}}\Big)^{1/3}\Big(\frac{\theta_{\mathrm{scatt}}}{\theta_{\mathrm{s}}}\Big)^{7/6}
\label{eq:mindref}
\end{equation}
where $r_{\mathrm{ref}}={r_{\mathrm{f}}}^2/r_{\mathrm{diff}}$ and
$\theta_{\mathrm{scatt}}=r_{\mathrm{ref}}/D$. The factor
$\theta_{\mathrm{s}}/\theta_{\mathrm{scatt}}$ can be replaced with 1 in the
case that $\theta_{\mathrm{s}}<\theta_{\mathrm{scatt}}$. Considering that 
$\theta_{\mathrm{scatt}}>\theta_{\mathrm{f}}$, we can assume this condition to
be true when dealing with interplanetary scintillation of compact radio
sources. Using Eq. \ref{eq:tsref} and \ref{eq:mindref}, we calculated the
values of $v$ and $r_{\mathrm{diff}}$ which are consistent with the observed
variability characteristics (a modulation index of a few percent and
$\tau_{\mathrm{c}} \sim 10^4$--$10^5$\,s). They are plotted in
Fig.\,\ref{fig:refpar}. The range of values which are consistent with the
solar wind speed are marked in black. It appears that in order to explain the
detected variability by means of refractive scintillation we must hypothesize
the existence of density inhomogeneities in the solar wind on a spatial scale of
tens of meters or less. Previous studies (see, e.g., Narayan et
al. \cite{Narayan1989} and Coles et al. \cite{Coles1991}), refer to an inner
scale of turbulence of the order of kilometers.   

\subsection{Scattering in the Earth's bow shock}

The sharp increase in the fractional modulation index at solar elongation
close to $30^\circ$ is compatible with another hypothesis, namely that the
variability increases due to propagation effects in the region where
the solar wind and the Earth's magnetosphere meet, i.e. the Earth's bow
shock. Here the solar wind speed drops considerably due to the interaction
with the Earth's magnetic field. It is worthwhile to investigate what values of $v$
and $r_{\mathrm{diff}}$ would be required to explain the variability
characteristics we found, considering a screen at a distance of $\sim10^8\,$m,
consistent with the approximative distance to the bow shock. This results in a Fresnel scale of
$\sim10^3\,$m and a Fresnel angle of the order of arcseconds, which implies that all
the sources can be treated as pointlike. In the case of weak scattering, a
variability time scale of $\sim 10^4\,$s leads to a velocity $v$ of the order
of a few m/s --- very low, even for the bow shock --- while from a modulation
index of the order of $1$\% we can deduce $r_{\mathrm{diff}}\sim10^5\,$m. For
refractive scintillation, instead, the constrains on $m_{\mathrm{i}}$ and
$\tau_{\mathrm{c}}$ would lead to $r_{\mathrm{diff}}\simeq$ m and, consequently,
$v\sim10^2\,$km/s. The value of $r_{\mathrm{diff}}$, in this case, is
much smaller than we would expect.

\subsection{Correlation between short- and long-term variability}

As shown above, both the hypotheses of weak and strong (refractive) scattering can
hold only assuming either a very slow moving component of the solar wind or
irregularities in the interplanetary plasma on very small size
scales. For typical parameters of the
interplanetary plasma, instead, IPS should cause variability on time scales
of seconds. Since the Urumqi observations are performed through sub-scans
which have a duration of about 30 seconds, an IPS effect would likely
appear as an additional noise superimposed to the Gaussian profile of
the flux density measurements. An increase in the noise caused by IPS
should translate into a larger uncertainty in the flux density
measurement. 

\begin{figure}
   \centering
   \includegraphics[width=0.96\columnwidth]{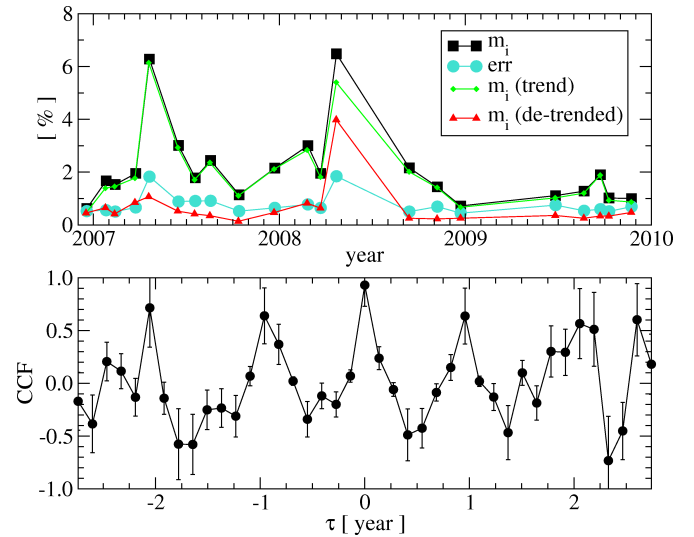}
      \caption{Upper panel: the modulation index (black dots) and the
        average flux density measurement errors (cyan dots) of
        AO 0235+164 plotted versus time. The errors are given in
        percentage with respect to the average flux density of the
        source during the $\sim3$ years of observation. The green
        squares and red triangles show the modulation index of the
        slow and of the fast variability component,
        respectively. Lower panel: the correlation function between
        the modulation index and the flux density measurement errors.
      }
              \label{fig:err}
\end{figure}

For all the observing sessions, we estimated the mean error on the flux
density measurements derived from the single sub-scans. The aim was to check if
the variations in the modulation index could be associated to similar 
variations in the flux density measurement error. Note that in the
error estimation there is a component which depends on the flux
density. Such a component can vary considerably for a strongly
variable source such as AO 0235+164 and must be removed. We used the
calibrators to derive the proportionality factor,
$\alpha_{\mathrm{n}}$, between the flux density ($S$) and the error
($\Delta S$). The flux-independent error for AO 0235+164 was
estimated as $(\Delta S)'=\Delta S - \alpha_{\mathrm{n}}\,S$. The
results are presented in the upper panel of Fig. {\ref{fig:err}}. The
cyan dots show the ratio between the errors and the average flux
density of AO 0235+164 calculated over the $\sim3$ years of observation,
expressed in percentage. This is meant to give an estimation of the
contribution of flux-independent error to the measurement
uncertainty. Such a contribution is small compared to the variability
on longer time scales, estimated through the modulation index (black
dots in Fig. {\ref{fig:err}}). 
However, a locally normalized discrete
correlation function (see Edelson \& Krolik \cite{Edelson1988}, Lehar
et al. \cite{Lehar1992})
between the two parameters shows a very high degree of correlation
(Fig. {\ref{fig:err}}, lower panel), higher than 0.9. Remarkable are
also the peaks of correlation for time lags of $\pm1$ and $\pm2$
years, which confirm the one-year periodical nature of the variations
in both the parameters.

Let us hypothesize that the
measurement error directly affects the modulation index by introducing
uncorrelated noise in the flux density estimation, due to the
uncertainty in the measurements. The periodic increase in the
modulation index would correspond to an equivalent increase in the
contribution of uncorrelated noise to the overall variability. This,
however, is not the case. Since uncorrelated noise has a time scale
which must be of the order of the average sampling of the light
curves, it can be isolated by separating the very fast variability
from the slow one throughout a de-trending procedure (see Villata et
al. \cite{Villata2002}). The variability
amplitude of the de-trended data (red triangles in
Fig. {\ref{fig:err}}, upper panel) provides us with an upper limit to
the contribution due to the measurement uncertainty. The comparison
between the modulation index of the de-trended light curves  and the
ones of the long-term trends (green squares in Fig. {\ref{fig:err}},
upper panel) demonstrates that the measurement uncertainty plays a
marginal role for the total variability, except maybe for the cases of
the December 2006 and April 2008 observing sessions. Once we excluded
that the measurement error directly affects the modulation index, we
can conclude that the correlation between the two is due to a common
origin of their variations.

\begin{figure}
   \centering
   \includegraphics[width=0.96\columnwidth]{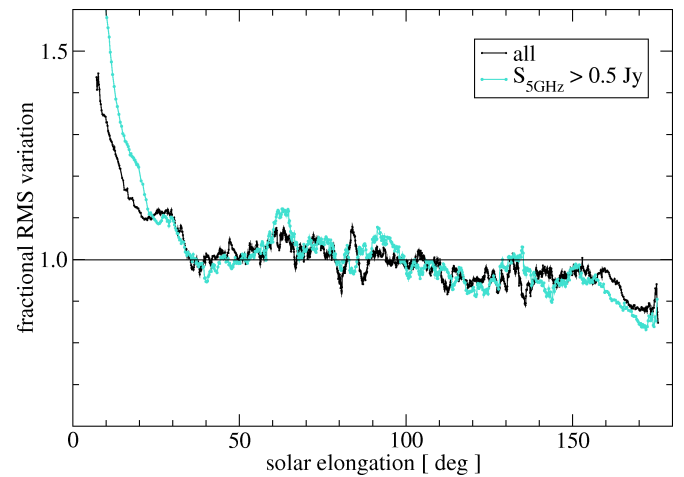}
      \caption{
        The 50-point running averages of the fractional RMS of the
        error-in-the-mean for all the MASIV sources (black line) and
        for a sub-sample of strong sources (cyan line). The solar
        elongation dependence is more pronounced in the latter ones.
      }
       \label{fig:rms}
\end{figure}

Concerning the MASIV data, possible variability on IPS time
  scales may be investigated by looking at the RMS error-in-the-mean
  for each scan. This is calculated from the scatter in the
  visibilities over all sub-integration times (typically 18 per
  one-minute scan, H. Bignall, priv. comm.) and all baselines
  (typically 10). The RMS error-in-the-mean for all the scans  were
  kindly provided by J. Lovell. Following the same approach as in
  Eq. \ref{eq:fm}, for each source we calculated the fractional RMS
  error-in-the-mean at each observing session, and we plotted it as a
  function of solar elongation. Its 50-point running average is shown
  in Fig. \ref{fig:rms} (black line). 
  The parameter follows a similar increasing trend as the modulation index,
  showing a prominent increase in
  the fractional variation for solar elongations $\epsilon$ below 35$^\circ$.  
  Remarkably, the solar elongation dependence of the RMS is more
  pronounced in the stronger MASIV sources 
  (S$_{\mathrm{5GHz}}>0.5$\,Jy, cyan line
  in the figure). 
  Likely, while for low flux density sources the RMS scatter on very short
  time scales is mainly due to thermal noise and confusion, for strong sources it is
  dominated by IPS. 

\begin{figure}
   \centering
   \includegraphics[width=0.96\columnwidth]{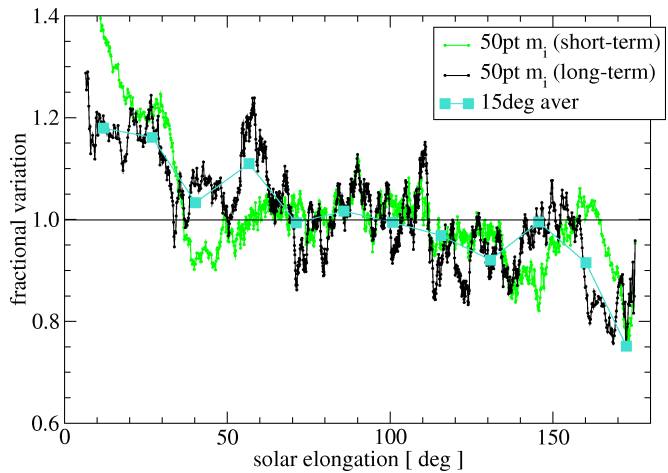}
      \caption{50-point running averages of the fractional modulation
        index of both the long- (black line) and the short-term
        variability (green line) in the MASIV data. The cyan squares show a
        15-degree average on the long-term data.}
       \label{fig:dtr-trn}
\end{figure}

For each MASIV light curve, we separated the long-term trend (which
carries information about the variability on time scales of one day or
more) from the fast variability component, calculating the modulation index of
both. We repeated the analysis presented in Fig. \ref{fig:fmall} for
the trend and the de-trended data separately. The
results, plotted in Fig. \ref{fig:dtr-trn}, show a strong solar-elongation
dependence in the short-term variability (green line). A similar trend, weaker
but still important, also appears in the long-term variability (black line), as
a 15-degree average (cyan squares) clearly shows. This suggests that
the influence of the Sun on the variability of compact sources goes
far beyond the typical IPS time scales.

The hypothesis that the variability of AO 0235+164 is related to IPS
 is strengthened by the results discussed above. They can be summarized
as follows: solar-elongation related variability on IPS time scales appears both in the Urumqi and in the VLA data. Looking at the Urumqi observations, we found evidence that this fast variability component correlates with the variability on typical IDV time scales.
 Still, to explain the phenomenon in terms of IPS we would need to postulate the existence of an \emph{atypical}, long-term manifestation of IPS. A possible
explanation would be that the variability on time scales of seconds do not
introduce a random error in the flux density measurement. We could
hypothesize that the scattering causes preferentially an
underestimation of the flux density of the source, by increasing
  the opacity of the scattering medium. In this case, a
stronger scattering --- due to a more turbulent medium in proximity of
the Sun --- would lead to lower values of the measured flux
density. Changes in the average conditions of the medium along the
line-of-sight, which could reasonably take place on time scales of
hours or more, or changes in the line-of-sight itself, would appear as a long-term trend in the light curve of
the source. A possible confirmation of this picture may come from the results of a study about extreme scattering events, reported by Lazio et al. (\cite{Lazio2001}). Using the Green Bank Interferometer, the authors monitored 149 compact radio sources with an average sampling of about one data-point every two days. They found evidence for a remarkable decrease (10-20\%) in the 2\,GHz flux density of 0952+179 in two separate occasions. Both events happened when the source was at low solar elongation. The time scales of the variations was of the order of weeks, which would be compatible with the long-term manifestation of IPS hypothesized above. At 8\,GHz frequency, however, the drop in flux density was not seen.

The existence of a correlation between slow and fast variability may also open the
way to a completely different explanation. Observations performed
close to the line-of-sight of the Sun may be strongly affected by
atmospheric effects (e.g. turbulence in the ionosphere). 
These are usually ignored because
of their short time scales, but if a connection between short and long
time scale variability could be hypothesized in the case of the
interplanetary medium, the same could also be valid for the
atmosphere.

\begin{figure}
   \centering
   \includegraphics[width=0.96\columnwidth]{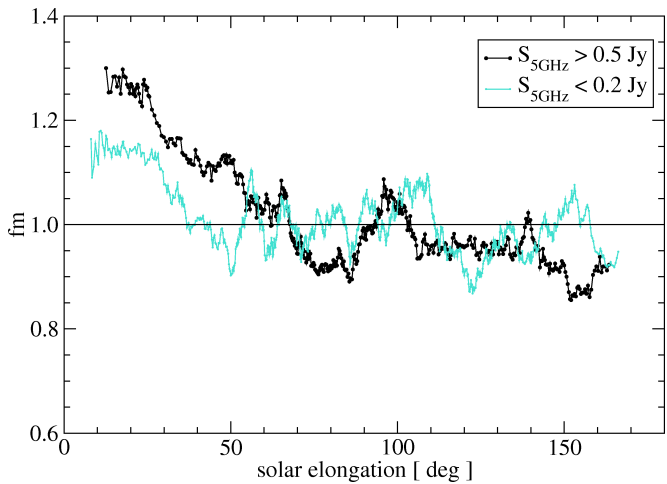}
     \caption{Results of a 40-point running average
        on two subsets of MASIV sources. The black curve concerns
        sources with S$_{5\mathrm{GHz}}>0.5\,$Jy, the cyan line
        sources with S$_{5\mathrm{GHz}}<0.2\,$Jy. For strong sources
        the solar-elongation dependence of the variability is much
        more prominent than for weak sources.
              }
              \label{fig:fmstrong}
\end{figure}

  In section \ref{comp}, we underlined that the  $\Delta
  m_{\mathrm{max}}$ value obtained for AO 0235+164 from all the Urumqi
  observations (i.e. including the two April sessions) is much higher
  than the one obtained from the MASIV data. This could be caused by
  changes occurring either in the source structure or in the
  scattering screen during the few years which separate the
  experiments; but it could also be caused by the difference between
  the facilities, in the sense that the Urumqi telescope may be more
  sensitive to the effect than the VLA. This may point towards
  a possible technical problem. An obvious candidate would be the
  effect of solar radiation into the sidelobes of the telescope or the
  receiver. There are facts, however, which 
  seem to rule out this hypothesis. If we repeat the analysis shown in
  Fig. \ref{fig:fmall} on sub-samples of strong and weak MASIV sources
  (with S$_{\mathrm{5GHz}}>0.5$\,Jy and $<0.2$\,Jy, respectively; see Fig. \ref{fig:fmstrong})
  we find that the former are considerably more affected than the
  latter ones. 
  This is consistent with the results concerning the RMS error-in-the-mean
  of the scans, discussed above. An additional
  contribution to the variability by the solar radiation 
  should have appeared more clearly in the light curves of weak
  sources. Furthermore, the solar radiation would be expected to introduce
  flux density fluctuations which are comparable for sources of
  similar brightness. Instead, the large variations observed in the
  modulation index of AO 0235+164 do not appear in CTA\,21, a
  steep-spectrum source which is only a few degrees away from AO 0235+164
  and is similarly bright. Still at a solar elongation of $\sim 20^\circ$, the
  modulation index of CTA\,21 is of the order of 0.5\%. 
    
  Another possible technical problem which could affect flux density
  measurements and depend on solar elongation is the pointing
  error. Reporting about the discovery of flickering in compact radio
  sources,  Heeschen (\cite{Heeschen1984}) specified that a
  considerable number of flux density measurements were excluded from
  the analysis because of the increase in the pointing errors at
  low solar elongations. We calculated the average pointing offsets
  for all the AO 0235+164 Urumqi observations. The April 2007 pointing
  offset turned out to be the second lowest in the time span between
  2005 and spring 2007, when the pointing model of the
  telescope was changed. Concerning April 2008, the pointing offset is
  lower than the average one  calculated between 2007 and 2010. This
  suggests that a pointing problem cannot account for the
  solar-elongation related variability observed in Urumqi. 

\subsection{The AO 0235+164 variability characteristics in April 2007 and 2008}

The large difference between the characteristic time scales of
AO 0235+164 in April 2007 and April 2008 deserves a final remark. While
in the first epoch most of the variability is due to a slow component
($\tau_{\mathrm{c}}\sim 2$ days), the April 2008 light curve is characterized
by a variability time scale which is comparable to the average sampling
($\tau_{\mathrm{c}}$ of the order of one hour, see
Fig.\,\ref{fig:lc}). As already mentioned, this may indicate that the
variability characteristics of the latter are dominated by the
uncertainty on the flux measurements. The high amplitude variations
observed in AO 0235+164 during the two epochs may be caused by effects
which are correlated, but are not the same. Further observations are
needed to understand at which solar elongation the fast variability
component starts to dominate over the slow one, if and how this can be
influenced by the activity of the Sun, and if the characteristics of
the observing facility play a role in it.

It cannot be excluded that also the conditions in the source itself
have to be taken into consideration. Differently from April 2008, in
April 2007 AO 0235+164 was in a flaring state (see Raiteri et
al.\,\cite{Raiteri2008}). The ejection of a strong and very compact
emitting component could have caused an additional variability
  contribution with characteristic time scale of $\sim 2$ days, due to
  ISS. The superposition of the variability contributions due to IPS
  and ISS could be the origin of the difference between the variability
  time scales in the two epochs.

\section{Summary}

This paper reports on the discovery of a seasonal cycle in the amplitude of
the variability in the IDV source AO 0235+164 during the
$\sim 3$ years of monitoring performed at the Urumqi Observatory. The
variability peaks at the time of minimum solar elongation of the source,
suggesting interplanetary scintillation as a possible cause of the
phenomenon. 

We performed a thorough investigation of the variability characteristics of a
sample of 475 sources, provided by the MASIV survey, with the aim of
establishing whether such a phenomenon generally enhances the
variability of compact radio sources. This study led to the conclusion that
solar-elongation related variability provides a significant contribution to
the total variability, especially for $\epsilon<30^\circ$. As expected, the
effect is most prominent for sources at low ecliptic latitude. We estimate
that for sources with $\beta\sim 0^\circ$ the variability increases by a
factor of two over the average source modulation index as they
pass from the maximum to the minimum solar elongation. The findings of
the present study may have important implications for future
large-area sky monitoring surveys (e.g. the proposed  ASKAP VAST
Survey; see Chatterjee et al. \cite{Chatterjee2010}).  

The nature of the variability is not yet completely understood. We took into
consideration the hypothesis of weak scattering and strong scattering from
refractive scintillation in the vicinity of the Sun as possible causes of the
phenomenon. In the first case, a very slow component of the solar wind (with speed
of the order of a $10$ meters per second) would be needed to explain the
observed variability characteristics. The refractive scintillation model,
instead, implies the existence of very small structures in the interplanetary
plasma. These conditions seem doubtful and do not find confirmation in previous studies in the
literature. We also took into consideration the hypothesis that the change of
variability is due to propagation effects in the Earth's bow shock. The
deduced values of $v$ and $r_{\mathrm{diff}}$ are not consistent with the
expected ones. 

We hypothesize that a change in the average parameters 
of the solar wind along the line-of-sight to the sources may provide a possible
explanation for the long time scale of the observed variability.

The strong correlation found in the Urumqi data between the measurement error on single
sub-scans and modulation index may be evidence of a link between IPS
(whose variability time scale is comparable with the duration of a
sub-scan) and the long-term variability. A similar kind of investigation
carried out on the MASIV data seems to support this idea.
However, alternative interpretations in terms of 
an \emph{indirect} effect of the Sun, atmospheric or even telescope-related 
effects cannot be excluded. To discriminate between the different hypotheses, 
further investigations are needed. 

\acknowledgements
{This paper made use of data obtained with the 25\,m Urumqi Observatory (UO)
  of the National Astronomical Observatories (NAOC) of the Chinese Academy of
  Sciences (CAS). Liu X. is supported by the National Natural Science
  Foundation of China under grant No.10773019 and 11073036 and the 973
  Program of China (2009CB824800). We would like to thank Peter M\"uller
  which provided the software for the processing of the Urumqi raw
  data. We would also like to thank Dr. Jim Lovell for
  providing us with the MASIV data and Dr. Hayley Bignall, Dr. Emmanouil Angelakis,
  Dr. Lars Fuhrmann and Dr. Axel Jessner for the stimulating discussions and
  suggestions.
}


\begin{thebibliography}{.99}

\bibitem[2003]{Bignall2003} Bignall, H.~E., Jauncey, D.~L., Lovell, J.~E.~J. et al.\ 
2003, \apj, 585, 653 

\bibitem[2010]{Chatterjee2010} Chatterjee, S., 
Murphy, T., \& VAST Collaboration 2010, Bulletin of the American Astronomical Society, 42, 515 

\bibitem[1991]{Coles1991} Coles, W.~A., Liu, W., Harmon, J.~K. \& Martin,
  C.~L.\ 1991, \jgr, 96, 1745

\bibitem[2000]{Dennett-Thorpe2000} Dennett-Thorpe, J., \& de Bruyn, A.~G.\ 2000, \apjl, 529, L65 

\bibitem[2002]{Dennett-Thorpe2002} Dennett-Thorpe, J., \& de Bruyn, A.~G.\ 2002, \nat, 415, 57 

\bibitem[1988]{Edelson1988} Edelson, R.~A., \& Krolik, J.~H.\ 1988, \apj, 333, 646 

\bibitem[2007]{Gabanyi2007} Gab{\'a}nyi, K.~{\'E}., Marchili, N., Krichbaum,
  T.~P. et al.\ 2007, \aap, 470, 83 

\bibitem[1984]{Heeschen1984} Heeschen, D.~S.\ 1984, \aj, 
89, 1111 

\bibitem[1987]{Heeschen1987} Heeschen, D.~S., 
Krichbaum, T., Schalinski, C.~J., \& Witzel, A.\ 1987, \aj, 94, 1493 

\bibitem[1997]{Kraus1997} Kraus, A.\ 1997, Ph.D.~Thesis, Bonn University,
  Germany 

\bibitem[1999]{Kraus1999} Kraus, A.,  Quirrenbach, A., Lobanov, A. P. et al.\ 1999, \aap, 344, 807 

\bibitem[2001]{Lazio2001} Lazio, T.~J.~W., Waltman, 
E.~B., Ghigo, F.~D., et al.\ 2001, \apjs, 136, 265 

\bibitem[2008]{Lazio2008} Lazio, T.~J.~W., Ojha, R., Fey, A.~L., et al.\ 2008,
  \apj, 672, 115 

\bibitem[1992]{Lehar1992} Lehar, J., Hewitt, J.~N., 
Burke, B.~F., \& Roberts, D.~H.\ 1992, \apj, 384, 453 

\bibitem[1976]{Lomb1976} Lomb, N.~R.\ 1976, \apss, 39, 447 

\bibitem[2003]{Lovell2003} Lovell, J.~E.~J., 
Jauncey, D.~L., Bignall, H.~E., et al.\ 2003, \aj, 126, 1699 

\bibitem[2008]{Lovell2008} Lovell, J.~E.~J., Rickett, B.~J., Macquart, J.-P. et 
al.\ 2008, \apj, 689, 108 

\bibitem[2008]{Marchili2008} Marchili, N., Krichbaum, T.~P., Liu, X., et al.\ 
2008, arXiv:0804.2787 

\bibitem[2009]{Marchili2009} Marchili N.\ 2009, Ph.D.~Thesis, Bonn University, Germany 

\bibitem[2010]{Marchili2010} Marchili, N., Mart\'i-Vidal, I., Brunthaler, A., et al.\ 2010, \aap, 509, A47 

\bibitem[1989]{Narayan1989} Narayan, R., Anantharamaiah, K.~R. \& Cornwell,
  T.~J.\ 1989, \mnras, 241, 403

\bibitem[1992]{Narayan1992}  Narayan, R.\ 1992, Royal 
Society of London Philosophical Transactions Series A, 341, 151 

\bibitem[2001]{Qian2001} Qian, S.-J. \& Zhang, X.-Z.\ 2001, \cjaa, 1, 133

\bibitem[1991]{Quirrenbach1991} Quirrenbach, A., Witzel, A., Wagner, S., et 
al.\ 1991, \apjl, 372, L71 

\bibitem[1992]{Quirrenbach1992} Quirrenbach, A., Witzel, A., Krichbaum, T.~P., et al.\ 1992, \aap, 258, 279 

\bibitem[2008]{Raiteri2008} Raiteri, C.~M., Villata, M., Larionov, V.~M., et al.\ 2008, \aap, 480, 339 

\bibitem[1971]{Readhead1971} Readhead, A.~C.~S.\ 1971, 
\mnras, 155, 185 

\bibitem[2001]{Rickett2001} Rickett, B.~J., Witzel, A., Kraus et al.\ 2001, \apjl, 550, L11  

\bibitem[2006]{Rickett2006} Rickett, B.~J., Lazio, 
T.~J.~W., \& Ghigo, F.~D.\ 2006, \apjs, 165, 439 

\bibitem[1982]{Scargle1982} Scargle, J.~D.\ 1982, \apj, 
263, 835 

\bibitem[2008]{Senkbeil2008} Senkbeil, C.~E., 
Ellingsen, S.~P., Lovell, J.~E.~J., et al.  \ 2008, \apjl, 672, L95 

\bibitem[1985]{Simonetti1985} Simonetti, J.~H., 
Cordes, J.~M., \& Heeschen, D.~S.\ 1985, \apj, 296, 46 

\bibitem[2006]{Sun2006} Sun, X.~H., Reich, W., Han, J.~L., et al.\ 2006, \aap, 447, 937 

\bibitem[2002]{Villata2002} Villata, M., Raiteri, C.~M., Kurtanidze,
  O.~M., et al.\ 2002, \aap, 390, 407 


\bibitem[1995]{Wagner1995} Wagner, S.~J., \& Witzel, A.\ 1995, \araa, 33, 163 

\bibitem[1986]{Witzel1986} Witzel, A., Heeschen, 
D.~S., Schalinski, C., 
\& Krichbaum, T.~P.\ 1986, Mitteilungen der Astronomischen Gesellschaft Hamburg, 65, 239 





\end{thebibliography}
\end{document}